\documentclass[10pt,a4paper,twoside]{article}
\usepackage{epsfig}
\usepackage{baltlat5}
\usepackage{wrapfig}
\begin{document}
\ \
\vspace{-0.5mm}

\setcounter{page}{1}
\vspace{-2mm}

\titlehead{Baltic Astronomy, vol.\ts 16, xxx--xxx, 2007.}

\titleb{MIXED CHEMISTRY PHENOMENON DURING LATE STAGES OF STELLAR EVOLUTION}

\begin{authorl}
\authorb{R. Szczerba, M.R. Schmidt, M. Pulecka}{1}
\end{authorl}

\begin{addressl}
\addressb{1}{Nicolaus Copernicus Astronomical Center,
ul. Rabia\'nska 8, 87-100 Toru\'n, Poland}
\end{addressl}

\submitb{Received 2006 October 15; revised ---}

\begin{summary}
We discuss phenomenon of simultaneous presence of O- and C-based material in 
surroundings of evolutionary advanced stars. We concentrate on silicate
carbon stars and present observations that directly confirm the
binary model scenario for them. We discuss also class of C-stars with OH emission 
detected, to which some [WR] planetary nebulae do belong. 
\end{summary}



\resthead{Mixed chemistry phenomenon}{R. Szczerba, M.R. Schmidt, M. Pulecka}

\sectionb{1}{INTRODUCTION}


During Asymptotic Giant Branch (AGB) phase of evolution stars with initial masses between 
0.8 and 8\,M$_{\odot}$ lose a significant amount of their initial mass by ejecting the 
matter into interstellar space with rates between 10$^{-7}$ and 10$^{-4}$ 
M$_{\odot}$\,yr$^{-1}$. 
The chemistry in the formed circumstellar envelopes is determined by the photospheric
C/O ratio and is O-based for n(O)$>$n(C) (usually less evolutionary advanced stars) 
and C-based when carbon abundance exceeds that of oxygen (evolved stars which experienced 
thermal pulses and dredged-up carbon to the surface). This dichotomy is a consequence of
CO molecule (very stable) formation which is so efficient that less abundant element
(C or O) is mostly used. Therefore, the detection of co-existence of O-rich and C-rich 
material in surroundings of evolved stars was (and still is) surprising and attracts a 
significant attention. Hereafter, we call this phenomenon a {\it mixed chemistry phenomenon}. 

Already, due to the IRAS observations it was realized that there is a group of carbon
stars which show typical for O-rich environment the 9.7 and 18\,$\mu$m amorphous silicate 
features (Little-Marenin 1986, Willems \& de Jong 1986). The Infrared Space Observatory 
(ISO) observations (Yamamura et al., 2000) showed that 9.7\,$\mu$m feature in one of such 
objects (V778 Cyg) is very stable and did not change during the last 15 years (the time 
spanned between IRAS and ISO observations). This put a very strong constraint on a model 
and evolutionary status of this class of objects with most likely explanation being a 
long-lived reservoir of O-rich material located inside or around a binary system. In this
review we discuss MERLIN interferometer observations of V778 Cyg which proved 
existence of such reservoir (disk) around companion of C-rich star. We note that the 
recent Spitzer Space Telescope (SST) data showed that the first extra-galactic 
silicate carbon star (IRAS\,04496$-$6859, Trams et al. 1999) is in fact a {\it normal} 
carbon star and do not show the 9.7\,$\mu$m dust emission (see Speck et al. 2006). 

There is another group of carbon stars suspected to have mixed chemistry. Namely, carbon 
stars with OH maser emission. Lewis (1992) listed a group of stars with SiC emission
seen in the IRAS Low 
Resolution Spectra (LRS) and OH maser emission detected. While most of these sources 
appeared to have wrong LRS classification the 3 C-stars with OH maser emission remained 
and Chen et al. (2001) added 6 more sources to this class. However, this class of 
{\it mixed chemistry} sources did not attract a significant attention (except of [WR] 
planetary nebulae -- see below), since OH emission is not well spatially resolved and this
group of sources may be result of spatial coincidence between OH maser emission from 
interstellar medium and location of C-star. For example, Szczerba et al. (2002) presented 
observational evidence that IRAS\,05373$-$0810 (C-star with OH maser emission) is 
a genuine carbon star and that OH maser and SiO thermal emission detected toward this 
star is not coming from its envelope, but from molecular clouds. Here we discuss a case of
another C-star with OH maser emission (IRAS~06238$+$0904) toward which we have detected, 
using IRAM radiotelescope, the SiO thermal emission coming from its envelope. Here, we 
present arguments that shock and Photon Dominated Region (PDR) chemistry allow to form a 
significant amount of SiO in C-rich environment.

One of the most important achievements of the ISO mission was detection of crystalline
silicates. Surprisingly, crystalline silicates were detected also in [WR] planetary 
nebulae, which have H-poor and C-rich central stars of WR-type\footnote{Note, that 
Zijlstra et al. (1991) detected OH maser emission from [WR] planetary nebula 
IRAS\,07027$-$7934. Therefore, at least this [WR] planetary nebula belongs also to the 
discussed above class of C-stars with OH maser emission.}.
[WR] planetary nebulae show at the same time presence of Polycyclic Aromatic Hydrocarbons
(PAHs) and crystalline silicates (Waters et al. 1998, Cohen et al. 1999). Scenarios 
proposed to explain simultaneous presence of PAHs and crystalline silicates include:
destruction of fossil comets orbiting the star, ejection of matter before star become
C-rich, formation of stable O-rich disk or torus around companion or system at some point
of binary evolution. Hajduk, Szczerba \& Gesicki (this Proceedings) present an attempt to 
determine spatial location of PAHs and crystalline silicates inside the [WR] planetary 
nebula M 2-43, by means of the radiative transfer modelling of ISO spectrum. They 
concluded that crystalline silicates have to be located at significant distance from the 
central star to avoid their emission at about 10 $\mu$m. We note also an attempt to find
precursors of [WR] planetary nebulae (C-rich stars with C- and O-rich material in their
circumstellar shells) among proto-planetary nebulae (Szczerba et al. 2003). The authors 
have argued that formation of crystalline silicates is necessary before proto-planetary 
nebula phase, while post-AGB star may be still O-rich and change to C-rich one
during the {\it fatal} thermal pulse. They indicated five proto-planetary nebulae as a 
possible precursors of [WR] planetary nebulae, including famous Red Rectangle, other 
C-rich source with crystalline silicates (IRAS 16279-4757), as well as three O-rich 
sources which show presence of crystalline silicates in their ISO spectra: AC Her, IRAS
18095$+$2704 and IRAS 19244$+$1115.

In this review we will not cover such cases as: NGC 6302 -- O-rich planetary nebula which 
show presence of crystalline silicates as well as PAHs (e.g. Kemper et al. 2002); HD 
233517 -- an evolved O-rich red giant with orbiting polycyclic aromatic hydrocarbons 
(Jura et al. 2006); IRAS 09425$-$6040 -- a carbon-rich AGB star with the highest abundance
of crystalline silicates detected up to now (Molster et al. 2001); IRC +10216 -- a well
known C-rich AGB star with water and OH maser lines detected (Melnick et al. 2001, 
Ford et al. 2003); and possibly some other spectacular sources which we, not intentionally, 
have overlooked.

\sectionb{2}{V778 CYG A SILICATE CARBON STAR}

To test the hypotheses related to the {\it mixed chemistry phenomenon} observed in silicate
carbon stars, we observed water masers towards V778\,Cyg at high angular resolution.  
Details of our observations and data analysis are presented by Szczerba et al. (2006), so 
here we repeat only some of the most important points and findings.

The observations were taken on 2001 October 12/13 under good weather
conditions, using five telescopes of MERLIN. The longest MERLIN baseline of 217\,km gave 
a fringe spacing of 12\,mas at 22.235080\,GHz. The bandwidth was 2\,MHz with 256 spectral 
channels per baseline providing a channel separation of 0.105\,km\,s$^{-1}$. 
The continuum calibrator sources were observed in 16\,MHz band with 16 channels.
The data were obtained in left and right circular polarisation and the velocities
were measured with respect to the local standard of rest.

We used the phase referencing method; 4\,min scans on V778\,Cyg were
interleaved by 2\,min scans on the source 2021+614 (at 3\fdg8 from the target) 
over 11.5\,h. The flux density of 2021+614 at K band of 1.48\,Jy was derived
from observation of 4C39.25. At the epoch of observation the flux density
of 3C39.25 was 7.5$\pm$0.3\,Jy (Terasranta 2002, private communication). This
source was also used for bandpass calibration.  

After initial calibration with the MERLIN software the data were processed
using the AIPS package. In order to derive phase and amplitude
corrections for atmospheric and instrumental effects the phase reference source
was mapped and self-calibrated. These corrections were applied to the target
visibility data. The absolute position of the brightest feature at 
$-$15.1\,km\,s$^{-1}$ was determined. The phase solutions for this feature were
obtained with self-calibration method and were then applied for the all channels.
The target was mapped and cleaned using a 12\,mas circular restoring beam. 
The map noise of $\sim$27\,mJy\,beam$^{-1}$ for $I$ Stokes parameter in a line-free 
channel was close to the predicted thermal noise level. 

In order to determine the position and the brightness of the maser components
two dimensional Gaussian components were fitted to the emission in channel maps.
The position uncertainty of this fitting depends on the signal to noise ratio
in the channel map and is lower than 1\,mas for about 80\%
of the maser components towards V778\,Cyg. The absolute position of the phase
reference source is known with an accuracy of $\sim$3\,mas.
The highest uncertainties in the absolute position of maser spots are due to 
tropospheric effects and errors in the telescope positions. The first effects, 
estimated by observing the phase rate on the point source 3C39.25, introduce 
the position error of $\sim$9\,mas, whereas uncertainties in telescope 
positions of 1$-$2\,cm cause an error of spot positions of $\sim$10\,mas. 
In order to check the position accuracy of maser spots we applied 
a reverse phase referencing scheme. The emission of 15 channels around 
the reference feature at $-$15.1\,km\,s$^{-1}$ was averaged and mapped. 
The map obtained was used as a model to self-calibrate the raw target data 
then these target solutions were applied to the raw data of 2021+614. 
The position of the reference source was shifted by $\sim$2\,mas with respect 
to the catalogue position. This indicates excellent phase connection when 
referencing 2021+614 to the set of the brightest maser spots. 
The above discussed factors imply the absolute position accuracy
of the maser source to be of order of $\sim$25\,mas.
\vbox{
\centerline{\psfig{figure=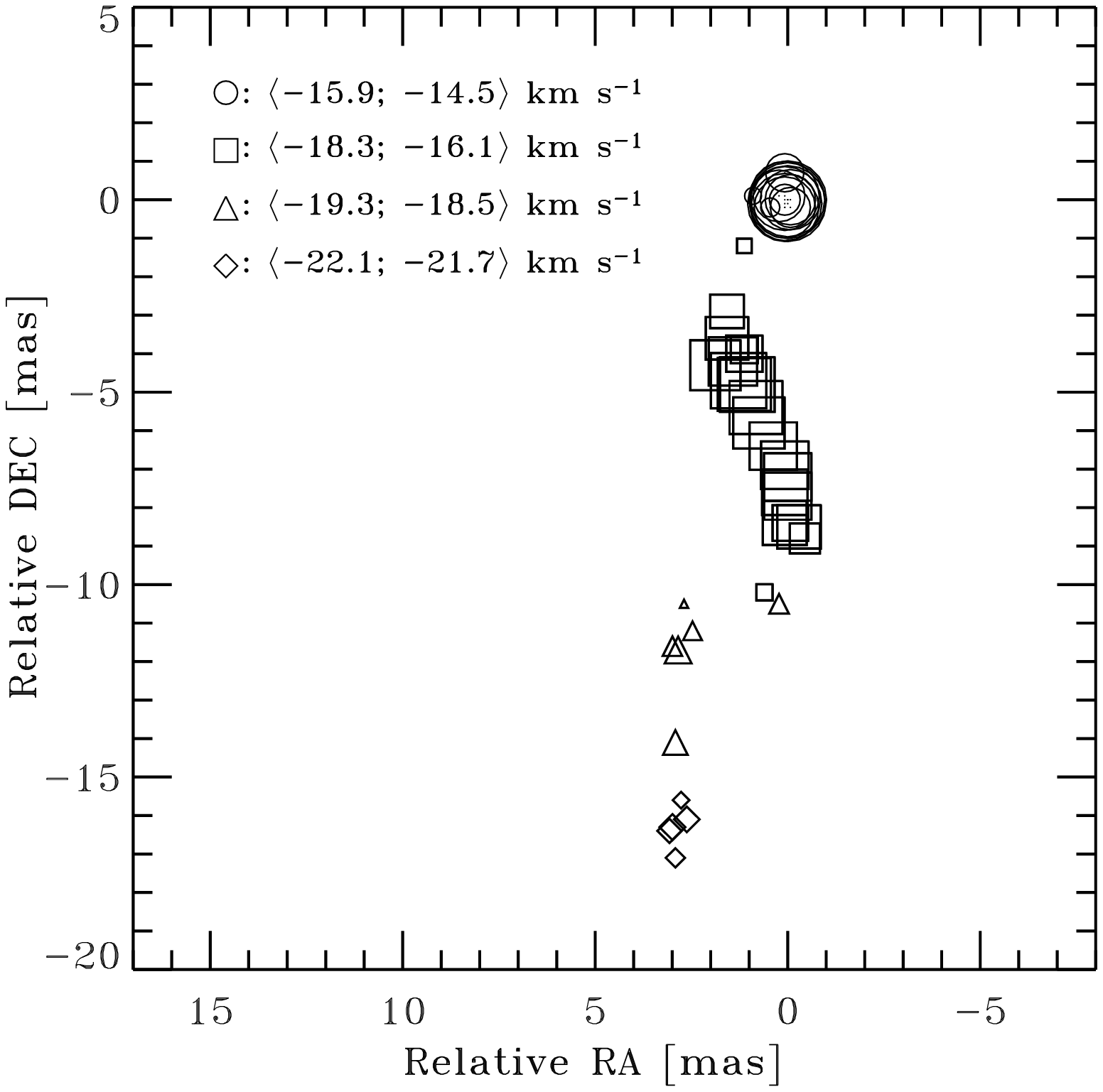,width=100mm,angle=0,clip=}}
\vspace{-2mm}
\captionc{1}{The absolute positions of the H$_2$O 22\,GHz maser
components towards V778\,Cyg relative to the reference feature 
at $-$15.1\,km\,s$^{-1}$ (RA(J2000) = 20$^{\rm h}$ 36$^{\rm m}$07\fs3833,
DE(J2000) = 60\degr05\arcmin26\farcs024). The symbols indicate the ranges 
of component velocities in km\,s$^{-1}$. 
The size of each symbol is proportional to the logarithm of peak brightness 
of the corresponding component.}
\vspace{2mm}
}
The maser emission brigther than 150\,mJ\,beam$^{-1}$ ($\sim5\sigma$)
was found in 51 spectral channels. In these channels
single and unresolved component only was detected. The overall structure
of the H$_2$O maser is shown in Fig. 1. The maser components form 
a distorted "S" like shape structure along a direction of position angle 
of about $-$10\degr. There is a clear velocity gradient along this structure
with weak south components blueshifted with respect to the brightest north
components. The angular extend of maser emisson is 18.5\,mas. 
The axis of alongation of the maser structure is fairly perpendicular
to the line towards the optical position of V778\,Cyg measured by Tycho2 (see Fig. 2).
Angular separation between the optical star and the maser reference component
is 0.192$\pm$0\farcs048.
\vbox{
\centerline{\psfig{figure=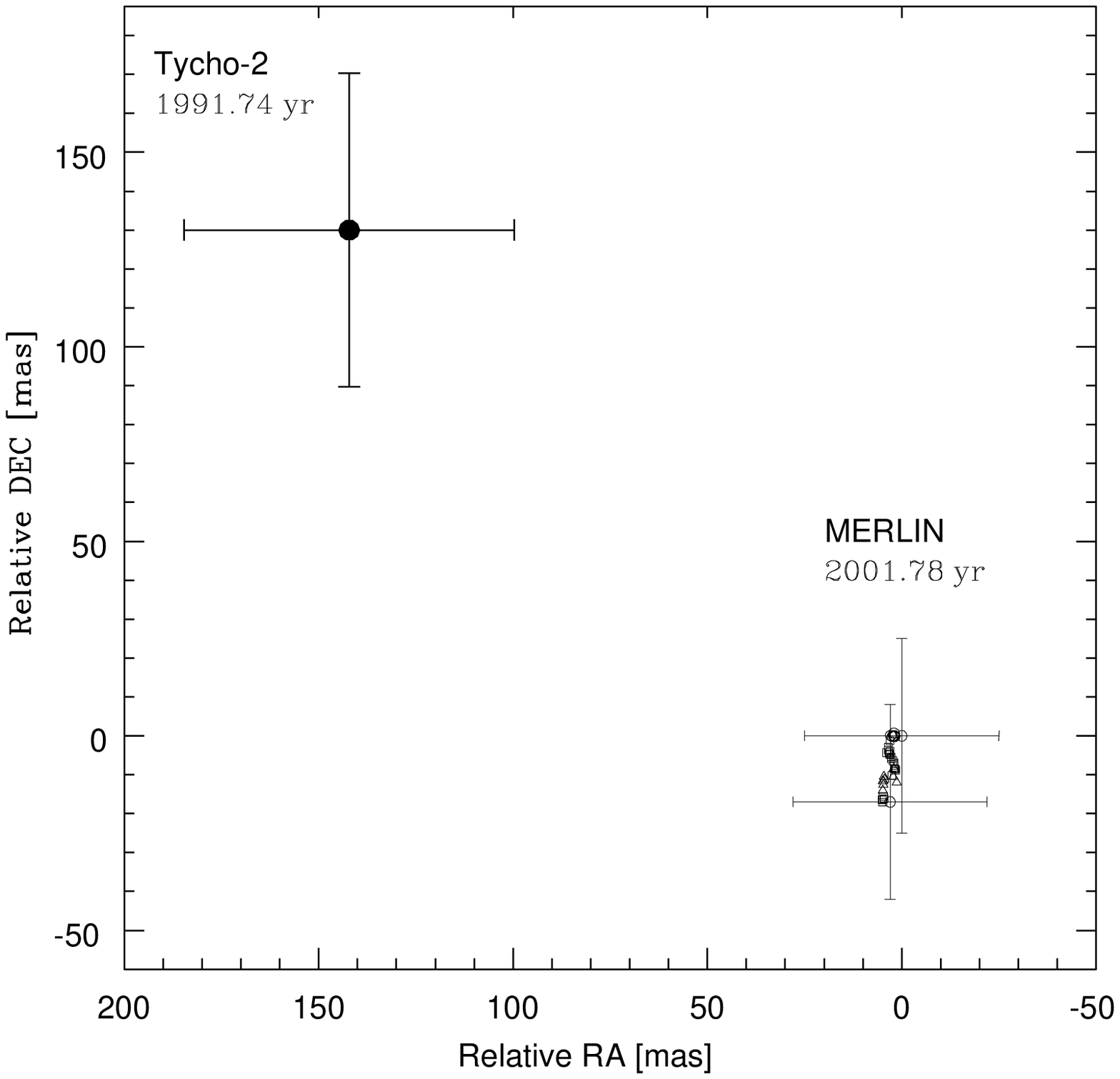,width=100mm,angle=0,clip=}}
\vspace{-25mm}
\captionc{2}{Comparison of th optical position of V778 Cyg as determined in the
Tycho-2 catalogue with the radio position of the H$_2$O 22 GHz maser 
components as obtained from the MERLIN measurements. The epochs of optical 
and radio observations differ by about 10 yrs.}
\vspace{2mm}
}
Szczerba et al. (2006) have argued that such separation cannot be explained by 
proper motion and instead provide direct observational evidence for the binary 
system model of Yamamura et al. (2000). They suggested that the observed water
maser components can be interpreted as an almost edge-on warped Keplerian disk
located around a companion object and tilted by no more 20\degr\ relative to the 
orbital plane. More detailed model of disk around companion in V778 Cyg system
is presented by Babkovskaia et al. (2006). Finally, note that recently 
Ohnaka et al. (2006) reported indirect detection of disk around another silicate 
carbon star (IRAS\,08002$-$38003). They argued that oxygen-rich material is stored in 
circumbinary disk surrounding the carbon-rich primary star and its putative low-luminosity
companion. These two findings may suggest that there are two different kinds of
silicate carbon stars: with circumbinary disk and disk around companion only.

\sectionb{3}{IRAS\,06238$+$0904 - AN OH MASER C-STAR OR GENUINE CARBON STAR?}

   Genuine carbon stars are formed during evolution on AGB. The star on that stage 
posses extended circumstellar envelope (CSE). In its inner part (near the 
photosphere) physical conditions (T$\sim$2500 K, $\rho\sim10^{14}$ 
cm$^{-3}$) make the material mainly molecular, with composition determined 
by the local thermodynamic equilibrium (LTE). In carbon CSE (C/O$>$1) after 
CO formation there is almost no oxygen left. However silicon monoxide (SiO) 
is observed in carbon stars. Recent observations (Sch\"{o}ier et al. 2006) 
show relatively high SiO fractional abundances ($1\times10^{-7} - 5 \times 10^{-5}$), 
while LTE models give $\sim 5 \times10^{-8}$ (Millar 2004). Therefore the 
non-equilibrium processes should be considered in modelling of circumstellar chemistry.

 In this review we focus on IRAS~06238+0904 -- an OH maser C-star (see Chen et al. 2001).
We first built model of carbon circumstellar envelope (CSE) and then computed radiative 
transfer in molecular rotational lines of HCN J=1-0, CS J=3-2, CS J=5-4 and SiO J=3-2,
detected by us with the IRAM radiotelescope. 

 Spectral energy distribution (SED) 
for IRAS~06238+0904 was modelled by means of the code and 
method described in Szczerba et al. (1997). 
The best fit (see Fig. 3) is obtained for the star's effective temperature
T$_*$=2500~[K], luminosity to distance ratio
$L/ d^2$=6500~[$L_{\sun} {\rm kpc}^{-2}$], mass loss rate $\dot{M}=
2 \times 10^{-5}$ [M$_{\sun}$\,yr$^{-1}$], dust temperature at the inner
boundary T$_{\rm dust}(R_{\rm in}^{\rm CSE}$)=900~[K], 
amorphous carbon (AC) and silicon carbide (SiC) to gas ratios: 
$\rho$(AC)/$\rho_{\rm gas}$=0.001, $\rho$(SiC)/$\rho_{\rm gas}$=0.00019.

\vbox{
\centerline{\psfig{figure=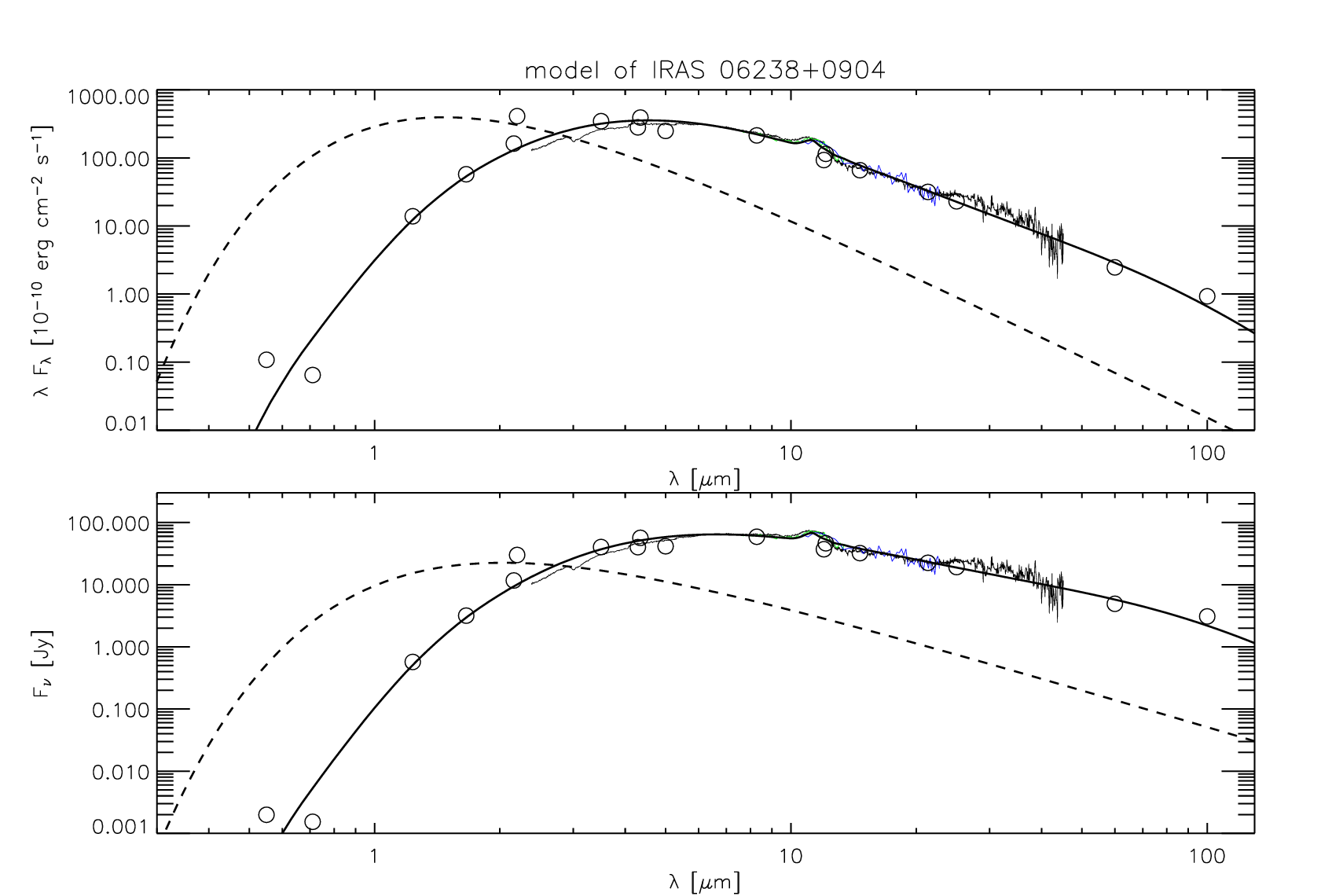,width=90mm,angle=90,clip=}}
\vspace{-2mm}
\captionc{3}{Spectral energy distribution for IRAS~06238+0904. See text for details 
 concerning assumed and estimated parameters.}
\vspace{2mm}
}

The chemical model is computed with the network based on RATE99 database
Le Teuff et al. (2000) composed of 343 species made of 10 elements. 
The gas temperature profile is approximated by the power law function
r$^{-1.8}$ established by iterations from the best fits to CS lines.
We assume solar gas composition with modifications of carbon
(C/O=1.5) and sulfur $\epsilon$(S)=6.71 abundances.
As initial concentrations we put LTE values of 23 important species, 
where SiO number density is equal 1$\times 10^{-8}$ [cm$^{-3}$]
The effect of dust formation is included by reduction of Si and C by 
amount locked up in SiC and amorphous carbon grains according to dusty model.
This results in decrease of silicon abundance to $\epsilon$(Si)=7.39 and 
decrease of C to O ratio to 1.3.

 The radiative transfer is computed in Sobolev approximation with molecular data
taken from the Leiden database (Sch\"{o}ier et al. 2005). Only 
interstellar radiation is taken into account as an important 
source of UV photons. Level populations of investigated molecules were 
computed for the assumed temperature and molecular densities resulting from chemical model.
The half-width main beam (HPBW) for SiO rotational transition J=3-2 (v=130 GHz) is equal 
to 18.9$\arcsec$, 16.7$\arcsec$ for CS(3-2), 10.0$\arcsec$ 
for CS(5-4) and 28.9$\arcsec$ for HCN(1-0) transition, in case of the IRAM telescope 
observations. The synthetic profile was computed for assumed distance to IRAS~06238+0904 
being equal 2.3 kpc.

 The observed and obtained molecular line profiles of SiO(3-2), CS(3-2) and CS(5-4)
are shown in 3 panels of Fig. 4. During line profiles modelling we included simple 
treatment of CO self-shielding based on Mamon et al. (1988). This process has considerable 
influence on all molecules, and is especially important for SiO. As one can see in Fig. 4, 
when self-shielding is not included (solid line) we can explain observed spectrum solely 
by the PDR chemistry. Around 1$\times 10^{16}$ [cm] we observed considerable reproduction
of SiO. On the other hand, inclusion of CO self-shielding prevents formation
of SiO (dashed line). Partial reproduction in PDR is still present. In both
cases the exchange reaction OH + Si $\rightarrow$ SiO + H is a main 
process responsible for formation of silicon monoxide. Exchanges between 
atomic oxygen and SiH, SiC, and HCSi molecules (O + SiH $\rightarrow$ SiO + H,  
O + HCSi $\rightarrow$ SiO + CH, and O + SiC $\rightarrow$ SiO + C) 
are also important.

Simulation of the shock passage (see Willacy \& Cherchneff 1998)
enlarge initial abundance of SiO, in comparison to the LTE value, 
for about one order. Profile obtain with abundance of this molecule 
increased by factor of 10 is shown as dashed-dotted line in Fig. 4.

\vbox{
\vspace{1mm}
\centerline{\psfig{figure=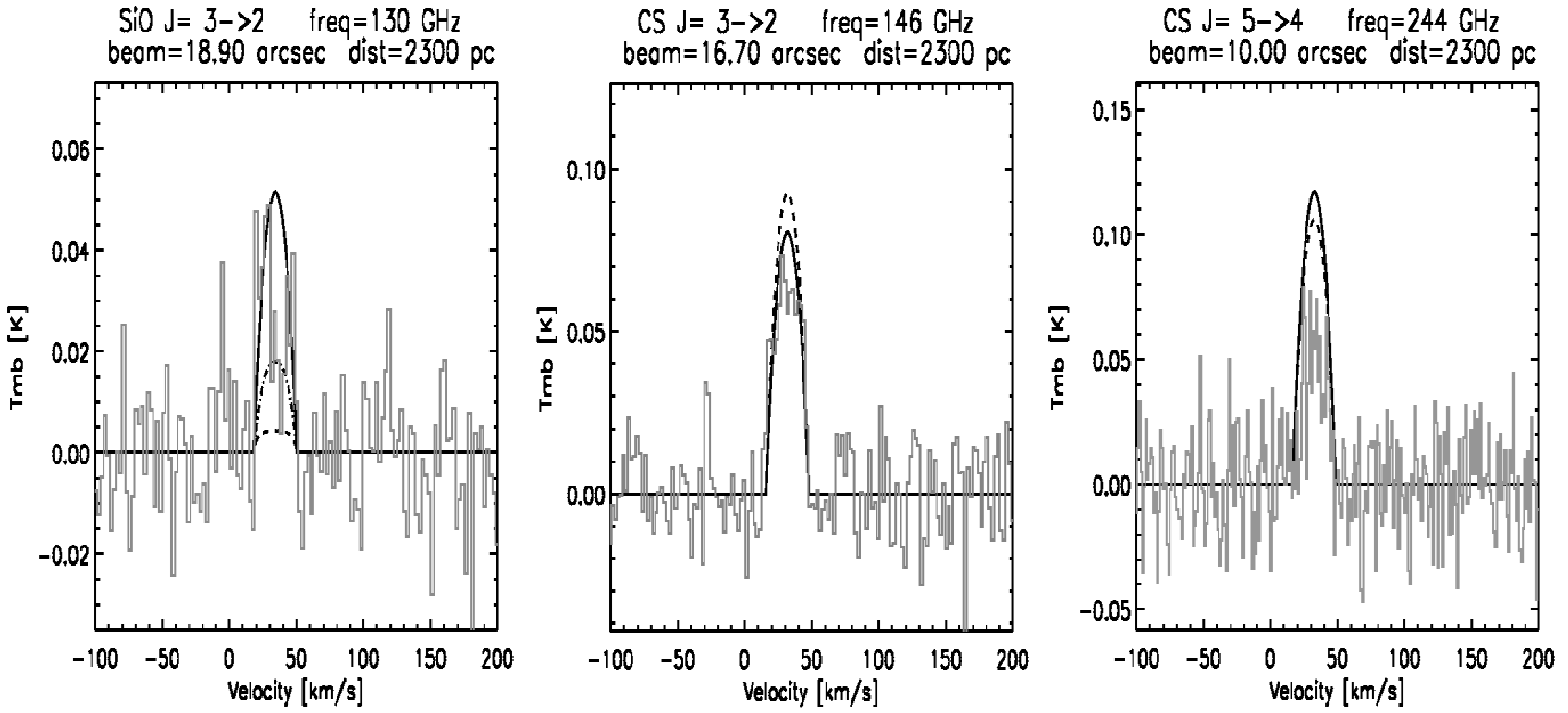,width=125mm,angle=0,clip=}}
\vspace{-2mm}
\captionc{4}{Observed and modelled molecular rotational lines without (solid line) and
with (dashed line) CO self-shielding. Dashed-dotted line in the left panel show results 
when the intial LTE abundance of SiO is increased ten times due to the shock passage.}
\vspace{1mm}
}
Therefore, we can conclude that IRAS~06238+0904 is a genuine C-star and no assumption
of {\it mixed chemistry} is necessary. Chemical reactions considered in network can 
reproduce O-bearing SiO molecule in C-rich environment if no CO self-shielding is 
considered. In presence of CO self-shielding the computed SiO emission is too low.
This may be improved, however, if we consider the effect of shock passage which can
increase the initial SiO abundance by order of magnitude as predicted by Willacy
\& Cherchneff (1998).

ACKNOWLEDGMENTS.
This work has been supported by grants 2.P03D.017.25 and 1.P03D.010.29
of the Polish State Committee for Scientific Research.


\References


\refb
Babkovskaia N., Poutanen J., Richards A. M. S., Szczerba R. 2006, MNRAS, 370, 1921

\refb 
Chen P. S., Szczerba R., Kwok S, Volk K. 2001, AA, 368, 1006

\refb
Cohen M., Barlow M. J., Sylvester R. J. et al. 1999, ApJ, 513, L135

\refb
Ford K. E. S., Neufeld D. A., Goldsmith P. F., Melnick G. J. 2003, ApJ, 589, 430

\refb
Jura M., Bohac C. J., Sargent B. et al. 2006, ApJ, 637, L45

\refb
Lewis B. M., 1992, ApJ, 396, 251

\refb
Le Teuff Y. H., Millar T. J., Markwick A. J. 2000, A\&AS, 146, 157 

\refb
Little-Marenin I. 1986, AA, 307, L15

\refb
Mamon G. A., Glassgold A. E., Huggins P. J., 1988, ApJ, 328

\refb
Melnick G. J., Neufeld D. A., Ford K. E. S. et al. 2001, Nature, 412, 160

\refb
Millar T. J. 2004, in AGB stars, eds. H. J. Habing, H. Olofsson, 247 

\refb
Molster F. J., Yamamura I., Waters L. B. F. M. et al. 2001, AA, 366, 923

\refb
Ohnaka K., Driebe T., Hoffman K.-H. et al. 2006, AA, 445, 1015

\refb
Sch\"{o}ier F., Olofsson H., Lundgren A. 2006, AA, 454, 247 

\refb
Sch\"{o}ier F., van der Tak F., van Dishoeck E., Black J. H. 2005, AA, 432, 369

\refb
Speck A., Cami J., Markwick-Kemper C. et al., 2006, ApJ, 650, 892

\refb
Szczerba R., Chen P. S., Szymczak M., Omont A. 2002, AA, 381, 491

\refb
Szczerba R., Omont A., Volk K. et al. 1997, AA.,317,859

\refb
Szczerba R. Stasi\'nska G., Si\'odmiak N., G\'orny S. K. 2003, in {\it Exploiting the ISO 
Data Archive. Infrared Astronomy in the Internet Age}, eds. C. Gry, S. Peschke, J. 
Matagne, P. Garcia-Lario, R. Lorente, A. Salama, ESA SP-511, 149 

\refb
Szczerba R., Szymczak M., Babkovskaia N. et al. 2006, AA, 452, 561

\refb
Trams N. R., van Loon J. Th., Zijlstra A. A. et al. 1999, AA, 344, L17

\refb 
Waters L. B. F. M., Beintema D. A., Zijlstra A. A. 1998, AA, 331, L61

\refb
Willacy K., Cherchneff I., 1998, AA, 330, 676 

\refb
Willems F. J., de Jong T. 1986, ApJ, 309, L39

\refb
Yamamura I., Dominik C., de Jong T. 2000, AA, 363, 629

\refb
Zijlstra A. A., Gaylard M. J., Te Lintel Hekkert P. et al. 1991, AA, 243, L9

\end{document}